\documentclass[aps,showpacs,twocolumn,eqsecnum,floatfix]{revtex4}

\usepackage{epsfig}
\usepackage{latexsym}

\begin{document}


\title{Regularization of spherically symmetric evolution codes in
numerical relativity}

\author{Miguel Alcubierre}
\email{malcubi@nuclecu.unam.mx}

\author{Jos\'e A. Gonz\'alez}
\email{cervera@nuclecu.unam.mx}

\affiliation{Instituto de Ciencias Nucleares, Universidad Nacional
Aut\'onoma de M\'exico, A.P. 70-543, M\'exico D.F. 04510, M\'exico.}


\date{\today}


\begin{abstract}
The lack of regularity of geometric variables at the origin is often a
source of serious problem for spherically symmetric evolution codes in
numerical relativity.  One usually deals with this by restricting the
gauge and solving the hamiltonian constraint for the metric.  Here we
present a generic algorithm for dealing with the regularization of the
origin that can be used directly on the evolution equations and that
allows very general gauge choices.  Our approach is similar in spirit
to the one introduced by Arbona and Bona for the particular case of
the Bona-Masso formulation.  However, our algorithm is more general
and can be used with a wide variety of evolution systems.
\end{abstract}




\maketitle


\section{Introduction}
\label{sec:introduction}

When developing spherically symmetric codes in numerical relativity,
the coordinate singularity at the origin can be a source of serious
problems caused by the lack of regularity of the geometric variables
there. The problem arises because of the presence of terms in the
evolution equations that go as $1/r$ near the origin.  Regularity of
the metric (essentially local flatness) guarantees the exact
cancellations of such terms at the origin, thus ensuring well-behaved
solutions.  This exact cancellation, however, though certainly true
for analytical solutions, usually fails to hold for numerical
solutions.  One then finds that the $1/r$ terms do not cancel and the
numerical solution becomes ill-behaved near $r=0$: it not only fails
to converge there, but can easily turn out to be violently unstable in
just a few timesteps.

The usual way to deal with this problem is to use the so-called {\em
areal} (or {\em radial}) gauge, where the radial coordinate $r$ is
chosen in such a way that the proper area of spheres of constant $r$
is always $4 \pi r^2$.  If, moreover, one also choses a vanishing
shift vector one ends up in the standard {\em polar/areal}
gauge~\cite{Bardeen83,Choptuik91}, for which the lapse is forced to
satisfy a certain ordinary differential equation in $r$.  The name
{\em polar} comes from the fact that for this gauge choice there is
only one non-zero component of the extrinsic curvature tensor, namely
$K_{rr}$~\cite{Bardeen83}.  In the polar/areal gauge the problem of
achieving the exact cancellation of the $1/r$ terms is reduced to
imposing the boundary condition $g_{rr}=1$ at $r=0$, which can be
easily done if one solves for $g_{rr}$ from the hamiltonian constraint
(which in this case is an ordinary differential equation in $r$) and
ignores its evolution equation.  If one does this in vacuum, one ends
up inevitably with Minkowski spacetime in the usual coordinates (one
can also recover Schwarzschild by working in isotropic coordinates and
factoring out the conformal factor analytically). Of course, in the
presence of matter, one can still have truly non-trivial dynamics.

The main drawback of the standard approach is that the gauge choice
has been completely exhausted.  In particular, the polar/areal gauge
can not penetrate apparent horizons, since inside an apparent horizon
it is impossible to keep the areas of spheres fixed without a
non-trivial shift vector.  The polar/areal gauge has, nevertheless,
been used successfully even in the study of critical collapse to a
black hole, where the presence of the black hole is identified by the
familiar ``collapse of the lapse'' even if no apparent horizon can be
found~\cite{Choptuik93}.

Still, one would like to have a way of dealing with the regularity
issue that allows more generic gauge choices to be made, either
because one is interested in studying the region inside an apparent
horizon, or because one wants to test interesting gauge conditions in
the simple case of spherical symmetry.  Because of this we have
developed a general regularization technique that can be used directly
on the Einstein evolution equations.

Our regularization method is similar in spirit, if not in detail, to
the one presented by Arbona and Bona in~\cite{Arbona98}.  The main
difference being that the approach of Arbona and Bona was tied to the
use of the Bona-Masso evolution
system~\cite{Bona89,Bona92,Bona93,Bona94b,Bona97a}, while our
algorithm is much more general.

A final point deserves notice.  Spherically symmetric evolution codes
that involve eternal black holes usually ignore the regularity
problem.  For example, one can excise the black hole interior and
eliminate $r=0$ from the numerical grid.  But even if one does not use
excision, for a black hole $r=0$ is not a regular point but rather a
compactification of the asymptotic infinity on the other side of the
Einstein-Rosen bridge.  This compactification introduces geometric
factors that compensate the $1/r$ terms making the equations regular
even at $r=0$.  This means that, contrary to what one would naively
expect, in spherical symmetry it is easier to evolve eternal black
holes (at least for some time) than it is to evolve regular
spacetimes.

This paper is organized as follows.  In Sec.~\ref{sec:regularity} we
discuss the regularity conditions that the metric functions must
satisfy at the origin of spherical coordinates, and we show which
terms need to be regularized in the Einstein equations.
Section~\ref{sec:algorithm} describes our regularization algorithm in
a generic way.  In Sec.~\ref{sec:examples} we present example of how
to regularize some specific formulations of the Einstein equations,
and show some numerical examples.  We conclude in
section~\ref{sec:discussion}.


\section{Regularity conditions}
\label{sec:regularity}

We start by writing the general form of the spatial metric in
spherical symmetry as
\begin{equation}
dl^2 = A(r,t) dr^2 + r^2 B(r,t) d \Omega^2 \; ,
\label{eq:metric}
\end{equation}
with $A$ and $B$ positive metric functions and $d \Omega^2$ the solid
angle element: $d \Omega^2 = d \theta^2 + \sin^2 \theta d \phi^2$.
Notice that we have already factored out the $r^2$ dependency of the
angular metric functions.  This has the advantage of making explicit
the dependency on $r$ of geometric quantities and makes the
regularization procedure easier.

As we will deal with the Einstein equations in first order form, we
will introduce the auxiliary quantities:
\begin{equation}
D_A := \partial_r \ln A \; , \qquad D_B := \partial_r \ln B \; .
\label{eq:defD}
\end{equation}
We will also work with the mixed components of the extrinsic
curvature: $K_A := K_r^r \, , K_B := K^\theta_\theta=K^\phi_\phi$.

There are in fact two different types of regularity conditions that
the variables $\{ A,B,D_A,D_B,K_A,K_B \}$ must satisfy at $r=0$.  The
first type of conditions are simple those imposed by the requirement
that the different variables should be well defined at the origin, and
imply the following behavior for small $r$:
\begin{eqnarray}
A &\sim& A^0 + {\cal O}(r^2) \; , \label{eq:regA} \\
B &\sim& B^0 + {\cal O}(r^2) \; , \label{eq:regb} \\D
_A &\sim& {\cal O}(r) \; , \label{eq:regDA} \\
D_B &\sim& {\cal O}(r) \; , \label{eq:regDB} \\
K_A &\sim& K_A^0 + {\cal O}(r^2) \; , \label{eq:regKA} \\
K_B &\sim& K_B^0 + {\cal O}(r^2) \; , \label{eq:regKB} 
\end{eqnarray}
with $\{ A^0,B^0,K_A^0,K_B^0 \}$ perhaps functions of time, but not of
$r$.  These regularity conditions are in fact quite easy to implement
numerically.  For example, one can use a finite differencing grid that
staggers the origin, and then obtain data on the fictitious point at
$r=-\Delta r /2$ by demanding for $\{ A,B,K_A,K_B \}$ to be even
functions at $r=0$ and for $\{ D_A,D_B \}$ to be odd.

It is the second type of regularity conditions that is more
troublesome.  In order to see the problem, we will first write the
Arnowitt-Deser-Misner (ADM) equations~\cite{Arnowitt62,York79} in
first order form for the case of spherical symmetry.  The evolution
equations are (in the case of zero shift)
\begin{eqnarray}
\partial_t A &=& -2 \alpha A K_A \; , \label{eq:Adot} \\
\partial_t B &=& -2 \alpha B K_B \; , \label{eq:Bdot} \\
\partial_t D_A &=& -2 \alpha [ K_A D_{\alpha} + \partial_r K_A ] \; ,
\label{eq:DAdot} \\
\partial_t D_B &=& -2 \alpha [ K_B D_{\alpha} + \partial_r K_B ] \; ,
\label{eq:DBdot} \\
\partial_t K_A &=& - \frac{\alpha}{A}  \left[ \partial_r ( D_{\alpha} 
+ D_B ) + D_{\alpha}^2 - \frac{D_{\alpha} D_A}{2} \nonumber \right. \\
&+& \frac{D_B^2}{2} - \frac{D_A D_B}{2}  - A K_A (K_A + 2 K_B) \nonumber \\
&-& \left. \frac{1}{r} (D_A - 2 D_B) \right] \; , \label{eq:KAdot} \\
\partial_t K_B &=& - \frac{\alpha}{2 A}  \left[ \partial_r D_B 
+ D_{\alpha} D_B + D_B^2 - \frac{D_A D_B}{2} \hspace{5mm} \nonumber \right. \\
&-& \frac{1}{r} (D_A - 2 D_{\alpha} - 4 D_B)
- \left. \frac{2 \left( A - B \right)}{r^2 B}  \right] \nonumber \\
&+& \alpha K_B (K_A + 2 K_B) \; \label{eq:KBdot}
\end{eqnarray}
where $\alpha$ is the lapse function and $D_{\alpha} := \partial_r \ln
\alpha$.  The hamiltonian and momentum constraints take the form
\begin{eqnarray}
\partial_r D_B &=& \frac{1}{r^2 B} \left( A - B \right)
+ A K_B \left( 2 K_A + K_B \right) \nonumber \\
&+& \frac{1}{r} \left( D_A - 3 D_B \right)
+ \frac{D_A D_B}{2} - \frac{3 D_B^2}{4} \; , \label{eq:ham} \\
\partial_r K_B &=& \left( K_A - K_B \right) \left[ \frac{1}{r}
+ \frac{D_B}{2} \right] \; , \label{eq:mom}
\end{eqnarray}

Since $\{D_\alpha,D_A,D_B\}$ go as $r$ near the origin, terms of the
type $D_{\{\alpha,A,B\}} /r$ are regular and represent no problem.
However, we see that both in the hamiltonian constraint and in the
evolution equation for $K_B$ there is a term of the form $(A-B)/r^2$,
while in the momentum constraint there is a term of the form $(K_A -
K_B)/r$.  Given the behavior of these variables near $r=0$ these terms
would seem to blow up at the origin.  The reason why this does not in
fact happen is that, near the origin, we must also ask for the extra
regularity conditions
\begin{equation}
A - B \sim {\cal O}(r^2) \; , \qquad K_A - K_B \sim {\cal O}(r^2) \; ,
\label{eq:regA-B}
\end{equation}
that is
\begin{equation}
A^0 = B^0 \; , \qquad K_A^0 = K_B^0 \; .
\end{equation}

It is not difficult to understand where these conditions come from.
They are just a consequence of the fact that space must remain locally
flat at $r=0$.  This local flatness condition implies that, near
$r=0$, it must be possible to write the metric as
\begin{equation}
{dl^2}_{R \sim 0} = dR^2 + R^2 d \Omega^2 \; ,
\end{equation}
with $R$ a radial coordinate that measures proper distance from the
origin.  A local transformation of coordinates from $R$ to $r$ then
takes the metric into the form
\begin{equation}
{dl^2}_{r \sim 0} = \left( \frac{dR}{dr} \right)^2_{r=0}
\left( dr^2 + r^2 d \Omega^2 \right) \; ,
\end{equation}
which implies that $A^0=B^0$ and, since this must hold for all time,
also that $K_A^0=K_B^0$.

It turns out that it is not trivial to implement numerically both the
symmetry regularity conditions~(\ref{eq:regA})-(\ref{eq:regKB}) and
the local flatness regularity conditions~(\ref{eq:regA-B}) at the same
time.  The reason for this is that at $r=0$ we now have three boundary
conditions for just two variables: both the derivatives of $A$ and $B$
must vanish, plus $A$ and $B$ must be equal to each other (and the
same thing must happen for $K_A$ and $K_B$).  The boundary conditions
for the exact equations are, of course, also over-determined, but in
that case the consistency of the equations implies that if they are
satisfied initially they remain satisfied for all time.  In the
numerical case, however, this is not true due to truncation errors,
and very rapidly (typically within one or two time steps) one of the
three boundary conditions fails to hold.  It is easy to convince
one-self that simply ignoring one condition and imposing the other two
does not work.  If we impose the zero derivative condition and ignore
the $A=B$ condition, then the $(A-B)/r^2$ term in the evolution
equations rapidly becomes singular.  On the other hand, if we impose
the $A=B$ condition and ignore one of the symmetry conditions, then we
introduce an inconsistency with the finite difference version of the
evolution equations, since for finite $\Delta r$ it is very difficult
to guarantee that the difference between $\partial_t K_A$ and
$\partial_t K_B$ approaches zero at the origin.  This inconsistency
then very rapidly causes large (and non-convergent) gradients to
develop near the origin.  In the following section we will introduce
an algorithm that successfully regularizes the numerical evolution
equations near \mbox{$r=0$}.

As a final comment, from the above equations we can also very easily
see why the polar/areal gauge has no serious regularity problem.  In
that gauge we have $B=1$ by construction.  If we now impose the
boundary condition $A(r=0)=1$, and solve for $A(r)$ by integrating the
hamiltonian constraint with $B=1$ and $D_B=0$ (ignoring the evolution
equations), then the $(A-B)/r^2$ term causes no trouble.


\section{Regularization algorithm}
\label{sec:algorithm}

In Ref.~\cite{Arbona98}, Arbona and Bona developed a regularization
technique for the spherically symmetric version of the Bona-Masso (BM)
evolution system~\cite{Bona89,Bona92,Bona93,Bona94b,Bona97a}.  Their
technique is based on redefining the auxiliary dynamical variable
$V_r$ that is part of the standard BM formulation in a way that allows
them to absorb the $(A-B)/r^2$ terms and reduces the regularization
problem to applying the correct boundary condition to $V_r$ at $r=0$.

Since we are interested in developing a generic regularization
technique, we will start from the ADM equations from the previous
section.  However, we will take the idea from the regularization
technique of Arbona and Bona of introducing an auxiliary variable
that will allow us to absorb the problematic terms.  Adding an
auxiliary variable seems to us to be the more straightforward way of
solving the problem of having over-determined boundary conditions: the
extra boundary condition will be imposed on the auxiliary variable.
We will then define the variable,
\begin{equation}
\lambda := \frac{1}{r} \left( 1 - \frac{A}{B} \right) \; .
\label{eq:lambda}
\end{equation}
Notice that, if the local flatness regularity
conditions~(\ref{eq:regA-B}) are satisfied, then the variable
$\lambda$ has the following behavior at the origin
\begin{equation}
\lambda \sim {\cal O}(r) \; ,
\label{eq:regl}
\end{equation}
which, as mentioned above, can easily be imposed numerically using a
grid that staggers the origin, and asking for $\lambda$ to be odd
across $r=0$.

The hamiltonian constraint now becomes
\begin{eqnarray}
\partial_r D_B &=& \frac{\lambda}{r}
+ A K_B \left( 2 K_A + K_B \right) \nonumber \\
&+& \frac{1}{r} \left( D_A - 3 D_B \right)
+ \frac{D_A D_B}{2} - \frac{3 D_B^2}{4} \; ,
\label{eq:regham}
\end{eqnarray}
and the evolution equation for $K_B$ becomes
\begin{eqnarray}
\partial_t K_B &=& - \frac{\alpha}{2 A}  \left[ \partial_r D_B 
+ D_{\alpha} D_B + D_B^2 - \frac{D_A D_B}{2} \hspace{5mm} \nonumber \right. \\
&-& \left. \frac{1}{r} (D_A - 2 D_{\alpha} - 4 D_B - 2 \lambda) \right]
\nonumber \\
&+& \alpha K_B (K_A + 2 K_B) \; .
\label{eq:regKBdot}
\end{eqnarray}
As mentioned, the problem terms have now been transformed into
$\lambda /r$, which will be well behaved as long as $\lambda$ is odd
at $r=0$.  The momentum constraint still has a term \mbox{$(K_A
-K_B)/r$}, but this should cause no trouble since it does not 
feed back into the evolution equations (one can always multiply the
momentum constraint with $r$ before evaluating it).  Of course,
multiples of the momentum constraint are typically added to the
evolution equations in order to build hyperbolic formulations, and we
will discuss how to deal with that term in such a case below.

There is one other ingredient that needs to be added: an evolution
equation for $\lambda$.  This can be obtained directly from its
definition:
\begin{equation}
\partial_t \lambda = \frac{2 \alpha A}{B} \left( \frac{K_A 
- K_B}{r} \right) \; .
\label{eq:lambdadot1}
\end{equation}
The last evolution equation clearly has the dangerous $(K_A - K_B)/r$
term, but this term can be removed with the help of the momentum
constraint~(\ref{eq:mom}) to find
\begin{equation}
\partial_t \lambda = \frac{2 \alpha A}{B} \left[ \partial_r K_B 
- \frac{D_B}{2} (K_A - K_B) \right] \; ,
\label{eq:lambdadot2}
\end{equation}
which is now regular at the origin.

The regularized first order ADM evolution equations are then
(\ref{eq:Adot})-(\ref{eq:KAdot}), with (\ref{eq:KBdot}) replaced by
(\ref{eq:regKBdot}), plus the evolution equation for $\lambda$ given
by (\ref{eq:lambdadot2}).

Having regularized the standard ADM equations, the question arises of
how to regularize alternative formulations where multiples of the
constraints can be added to the evolution equations in a number of
ways.  Adding multiples of the hamiltonian constraint represents no
problem, as the introduction of the variable $\lambda$ already
regularized this constraint, as seen in~(\ref{eq:regham}).  The
momentum constraint, however, is not regularized as it still includes
the term $(K_A - K_B)/r$.  One could try to play the same game as
before and introduce yet another variable to absorb this term.
However, we will now show that this is not really necessary.

Let us then consider some arbitrary first order formulation of the
Einstein evolution equations in spherical symmetry that has the
generic form
\begin{eqnarray}
\partial_t u_i &=& q_i(u,v) \; , \\
\partial_t v_i &=& M^j_i(u) \; \partial_r v_j + p_i(u,v) \; ,
\end{eqnarray}
where $u=(A,B,\lambda)$ and $v=(D_A,D_B,K_A,K_B)$.  The source terms
$q$ and $p$ are assumed not to depend on derivatives of any of the
fields.  The formulation might be hyperbolic or not, depending on the
characteristic structure of the matrix $M$.  We will assume that one
has arrived at such a formulation by adding multiples of the
hamiltonian and momentum constraints to the evolution equations for
the $v$'s.  This means that one can expect that the source terms $p_i$
will in general contain terms proportional to $(K_A - K_B)/r$.  We will
then rewrite the evolution equations for the $v_i$ as
\begin{equation}
\partial_t v_i = M^j_i(u) \; \partial_r v_j + p'_i(u,v)
+ \frac{f_i(u)}{r} (K_A - K_B) \; .
\end{equation}
Here we are assuming that the coefficient $f_i(u)$ of the $(K_A -
K_B)/r$ terms depends of the $u$'s, but not on the $v$'s, which will
typically be the case.  Using now equation~(\ref{eq:lambdadot1}) we
find
\begin{equation}
\partial_t v_i = M^j_i(u) \; \partial_r v_j + p'_i(u,v)
+ \frac{f_i(u) B}{2 \alpha A}
\; \partial_t \lambda \; ,
\end{equation}
which implies
\begin{eqnarray}
\partial_t \left( v_i - \frac{f_i(u) B}{2 \alpha A}
\; \lambda \right) &=& M^j_i(u) \; \partial_r v_j + p'_i(u,v) \nonumber \\
&-& \lambda \; \partial_t \left( \frac{f_i(u) B}{2 \alpha A} \right) \; .
\end{eqnarray}
If we now define
\begin{equation}
v'_i := v_i - \frac{f_i(u) B}{2 \alpha A} \; \lambda \; ,
\end{equation}
we can transform
the last equation into
\begin{equation}
\partial_t v'_i = M^j_i(u) \; \partial_r v_j + p'_i(u,v)
- \lambda \; F^t_i(u,v) \; ,
\end{equation}
with $F^t_i(u,v) = \partial_t \left( f_i(u) B / 2 \alpha A \right)$.
Notice that $F^t_i(u,v)$ so defined will involve no spatial
derivatives of $u$'s or $v$'s.  The final step is to substitute the
spatial derivative of $v_j$ for that of $v'_j$ to find
\begin{eqnarray}
\partial_t v'_i &=& M^j_i(u) \; \partial_r v'_j + p'_i(u,v)
- \lambda \; F^t_i(u,v) \nonumber \\
&+& \partial_r \left( \frac{f_i(u) B}{2 \alpha A}\; \lambda \right)\nonumber \\
&=& M^j_i(u) \; \partial_r v'_j + p'_i(u,v)
+ \lambda \left( F^r_i(u,v) - F^t_i(u,v) \right) \nonumber \\
&+& \left( \frac{f_i(u) B}{2 \alpha A} \right) \; \partial_r \lambda \; ,
\end{eqnarray}
with $F^r_i(u,v) = \partial_r \left( f_i(u) B / 2 \alpha A \right)$.  
Using now the fact that
\begin{equation}
\partial_r \lambda = -\frac{1}{r} \left[ \lambda
+ \frac{A}{B} \left( D_A - D_B \right)
\right] \; ,
\end{equation}
we finally find
\begin{eqnarray}
\partial_t v'_i &=& M^j_i(u) \; \partial_r v'_j + p'_i(u,v)
+ \lambda \left( F^r_i(u,v) - F^t_i(u,v) \right) \nonumber \\
&-& \frac{f_i(u) B}{2 \alpha A r} \left[ \lambda
+ \frac{A}{B} \left( D_A - D_B \right)
\right] \; .
\end{eqnarray}
This last equation is now regular, and has precisely the same
characteristic structure as the original system.  What we have done is
transform the original evolution equations for the $v_i$ variables
into evolution equations for the new $v'_i$ variables for which the
principal part terms are the same and the source terms are regular.
Notice that typically only some of the $f_i(u)$ will be different
from zero, so one does not need to transform all variables.

In the following section we will consider examples of how to regularize
some specific systems of evolution equations.


\section{Examples}
\label{sec:examples}

The regularized first order ADM equations where derived in the last
section.  Here we will examples of numerical evolutions using two
different regularized systems.

In the numerical simulations we will take as initial data Minkowski
spacetime in the usual coordinates, so that:
\begin{eqnarray} 
A &=& B \,\,\,\, = 1 \; , \\
D_A &=& D_B = 0 \; , \\
K_A &=& K_B = 0 \; .
\end{eqnarray}
In order to have a non-trivial evolution, we will chose an
initial lapse profile of the form:
\begin{equation} 
\alpha(t=0) = 1 + r^2 \hat{C} e^{-\frac{(r-\hat{r})^2}{\hat{\sigma}^2}} \; ,
\end{equation}
that is, we add a small gaussian contribution to the initial Minkowski
lapse.  We will then evolve the lapse using a Bona-Masso (BM)
slicing condition~\cite{Bona94b}, so the evolution equation for the
lapse will be:
\begin{equation}
\partial_t \alpha = - \alpha^2 f(\alpha) (K_A + 2 K_B) \; .
\end{equation}

In the simulations shown below we have taken the gaussian parameters
to be: $\hat{C} = 0.001, \hat{r} = 5.0$, $\hat{\sigma} = 1.0$.  We
have also restricted ourselves to harmonic slicing, that is,
$f(\alpha)=1$.

\subsection{System I}

As a first example we will build a hyperbolic system starting from the
ADM equations and the BM slicing condition.  We construct this system
by using the Hamiltonian and Momentum constraints to remove the terms
proportional to $\partial_r D_B$ and $\partial_r K_B$ from the
evolution equations of $K_A$ and $D_{\alpha}$ respectively.  The
resulting system can be easily shown to be strongly hyperbolic.

\begin{figure}
\vspace{5mm}
\epsfig{file=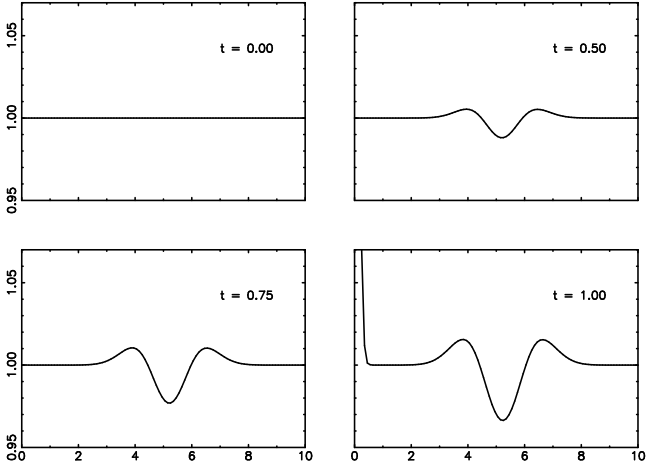,width=6.1cm,angle=270}
\caption{System I not regularized.  The plots show the evolution of
the metric function $A$ at different times.  Note the spike at $r=0$
for $t = 1$.}
\label{fig:A_no_reg} 
\end{figure}

\begin{figure}
\vspace{5mm}
\epsfig{file=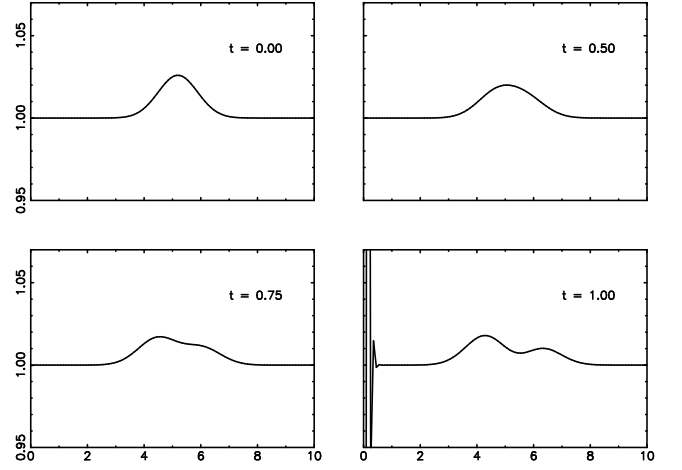,width=6.1cm,angle=270}
\caption{System I not regularized.  The plots show the evolution of
the lapse function $\alpha$ at different times. Note the spike at
$r=0$ for $t = 1$.}
\label{fig:alpha_no_reg}
\end{figure}

Figures~{\ref{fig:A_no_reg}} and~{\ref{fig:alpha_no_reg}} show the
evolution of the radial metric $A$ and lapse function $\alpha$ using
the system desribed above, with no regularization.  Note that both
plots show a spike at $r=0$ for times $t \approx 1$.

Next we look at the regularized case.  As described above, we first
introduce the auxilliary variable
\begin{equation}
\lambda = \frac{1}{r} \left( 1 - \frac{A}{B} \right) \; .
\end{equation}
Also, since we have used the momentum constraint to modify the evolution
equation for $D_{\alpha}$, we will need to replace this variable with
\begin{equation}
U_{\alpha} := D_{\alpha} + \frac{B \lambda}{A}  \; .
\end{equation}
The set of variables to be evolved is then: 
\begin{equation}
\{ \alpha, A, B, U_{\alpha}, D_A, D_B, K_A, K_B, \lambda \} \; .
\end{equation}
The evolution equations for these variables take the form:
\begin{eqnarray}
\partial_t \alpha &=& - \alpha^2 f ( K_A + 2 K_B) \; , \\
\partial_t A &=& - 2 \alpha A K_A \; , \\
\partial_t B &=& - 2 \alpha B K_B \; , \\
\partial_t U_{\alpha} &=&  - \alpha f \; \partial_r K_A + \alpha ( K_A + 2 K_B )
\left[ \frac{f^2 B \lambda}{A} \right. \nonumber \\
&-& \left. \rule{0mm}{5mm} U_{\alpha} \left( f+\alpha f'\right) \right]
+ \alpha f (K_B - K_A) \nonumber \\
&& \left( D_B -
\frac{2 B \lambda}{A} \right)\; , \\
\partial_t D_A &=& - 2 \alpha \left[ \partial_r K_A + K_A \left( U_{\alpha}
- \frac{f B \lambda}{A} \right) \right]   \; , \\
\partial_t D_B &=& - 2 \alpha \left[ \partial_r K_B + K_B \left( U_{\alpha}
- \frac{f B \lambda}{A} \right) \right] \; , \\
\partial_t K_A &=& - \frac{\alpha}{A} \partial_r U_{\alpha} - \frac{\alpha}{A}
\bigg\{ \left( U_{\alpha} - \frac{f B \lambda}{A} \right) \left[ U_{\alpha}
- \frac{D_A}{2} \right. \nonumber \\
&-& \left. \frac{B \lambda}{A} (f + \alpha f') \right] +
\frac{f B \lambda}{A} (D_A - D_B) - \frac{D_B^2}{4} \nonumber \\
&+& A \left( K_B^2 - K_A^2 \right) - \frac{1}{r}
\left[ \rule{0mm}{5mm} \lambda + D_B \right. \nonumber \\
&-& \left. f \left( \frac{B \lambda}{A} + D_A - D_B \right) \right] \bigg\}  \; , \\
\partial_t K_B &=& - \frac{\alpha}{2A} \partial_r D_B + \frac{\alpha}{2A}
\bigg\{ - \left( U_{\alpha} - \frac{f B \lambda}{A} \right) D_B - D_B^2
\nonumber \\
&+& \frac{D_A D_B}{2} + 2 A K_B (K_A + 2 K_B) + \frac{1}{r}
\left[ \rule{0mm}{5mm} D_A - 2 U_{\alpha} \right. \nonumber \\
&+& \left. \frac{2 f B \lambda}{A} - 4 D_B - 2 \lambda
\right] \bigg\} \; , \\
\partial_t \lambda &=& \frac{2 \alpha A}{B} \; \partial_r K_B
+ \frac{\alpha A D_B}{B} (K_B - K_A)  \; .
\end{eqnarray}
with $f':= df/d \alpha$.

\begin{figure}
\vspace{5mm}
\epsfig{file=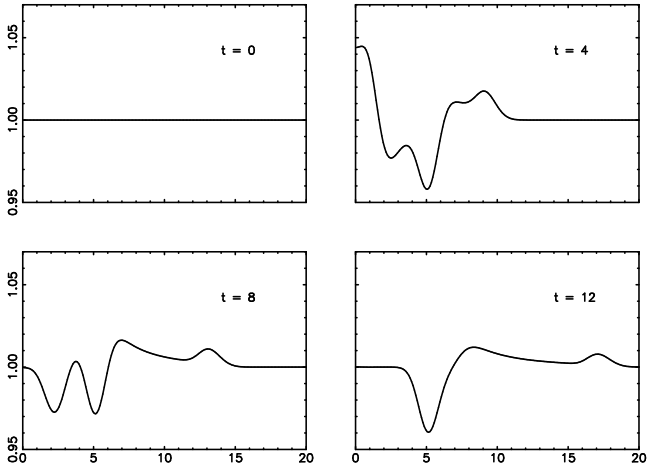,width=6.1cm,angle=270}
\caption{System I regularized.  The plots show the evolution of the
metric function $A$ at different times.}
\label{fig:A_reg}
\end{figure}

\begin{figure}
\vspace{5mm}
\epsfig{file=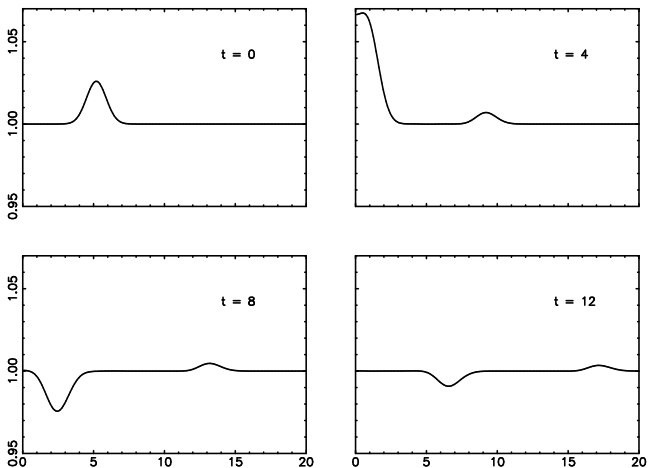,width=6.1cm,angle=270}
\caption{System I regularized.  The plots show the evolution of the
lapse function $\alpha$ at different times.}
\label{fig:alpha_reg}
\end{figure}

Figures~{\ref{fig:A_reg}} and~{\ref{fig:alpha_reg}} shown again the
evolution of $A$ and $\alpha$ for the regularized case.  The system
can now be evolved for very long times and the origin remains well
behaved. The plots show the evolution up to $t = 12$.

\subsection{System II}

Our second example is again a hyperbolic system built from the ADM
equations and the BM slicing condition, but now instead of using $D_A$
and $K_A$ as fundamental variables, we have used $\widetilde{D} = D_A
- 2 D_B$ and $K \equiv \rm{tr} K := K_A + 2 K_B$.  In order to make
the system hyperbolic we remove the terms proportional to $\partial_r
D_B$ and $\partial_r K_B$ from the evolution equations of $K$ and
$\widetilde{D}$ using the constraints. We then obtain a new strongly
hyperbolic system different from System I above.

For the regularization procedure we introduce again the variable
$\lambda$.  Since in this case we have used the momentum constraint to
modify the evolution equation for $\widetilde{D}$, we need to replace
this variable with
\begin{equation}
\widetilde{U} := \widetilde{D} - \frac{4 B \lambda}{A} \; .
\end{equation}
The final set of dynamical variables is then
\begin{equation}
\{ \alpha, A, B, D_{\alpha}, \widetilde{U}, D_B,  K, K_B, \lambda \} \; ,
\end{equation}
and their evolution equations are:
\begin{eqnarray}
\partial_t \alpha &=& - \alpha^2 f K \; , \\
\partial_t A &=& 2 \alpha A ( 2 K_B - K) \; , \\
\partial_t B &=& - 2 \alpha B K_B \; , \\
\partial_t D_{\alpha} &=&  - \partial_r ( \alpha f K ) , \\
\partial_t \widetilde{U} &=& - 2 \partial_r ( \alpha K ) + 4 \alpha D_B 
(K - 3 K_B) \nonumber \\ &+& 8 \alpha \left[ D_{\alpha} K_B 
+ \frac{B \lambda}{A} (3 K_B- K) \right] \; , \\
\partial_t D_B &=& - 2 \partial_r ( \alpha K_B ) , \\
\partial_t K &=& \alpha \left[ - 4 K K_B + 6 K_B^2 - \frac{2 D_{\alpha}}{A r} 
+ K^2  \right. \nonumber \\
&+& \left. \frac{D_{\alpha}}{2 A} \left( \widetilde{U} + \frac{4 \lambda B}{A} 
\right) - \frac{D_{\alpha}^2}{A} - \frac{\partial_r 
D_{\alpha}}{A} \right] \; , \\
\partial_t K_B &=& \frac{\alpha}{Ar} \left[ \frac{\widetilde U}{2} 
+ \frac{2 \lambda B}{A} - D_B - \lambda - D_{\alpha} \right] \nonumber \\
&+& \frac{\alpha}{A} \left[ -\frac{D_{\alpha} D_B}{2}  - \frac{\partial_r D_B}{2}
\right. \nonumber \\
&+& \left. \frac{D_B}{4}\left(\widetilde{U} 
+ \frac{4 \lambda B}{A} \right) + A K K_B \right] , \\
\partial_t \lambda &=& \frac{2 \alpha A}{B} \left[ \partial_r K_B 
- \frac{D_B}{2} (K - 3 K_B)\right]  \; .
\end{eqnarray}

In this case there is no need to show any plots, as the numerical
evolution behaves exactly in the way it did for System I, with the
origin remaining well behaved.


\section{Discussion}
\label{sec:discussion}

Lack of regularity of geometric variables at the origin is often a
problem for spherically symmetric evolution codes in numerical
relativity.  We have shown that the problem can be traced to the
existence of two types of regularity conditions at the origin.  In the
first place, there are regularity conditions that guarantee that the
variables are well defined at the origin.  These conditions can be
written down as a series of symmetry conditions at the origin for the
different variables, and can be easily enforced in numerical
simulations.  However, there also exist regularity conditions related
to the condition that spacetime must be locally flat at the origin.
Together, all these regularity conditions imply that we have more
conditions to satisfy at $r=0$ than dynamical variables, which means
that numerically some regularity conditions will inevitably be
violated.  We have presented a generic regularization algorithm that
is based on the introduction of an auxiliary variable that absorbs the
problematic terms and on which we can impose the extra boundary
conditions at $r=0$.  Our algorithm is similar in spirit, if not in
detail, to an algorithm presented by Arbona and Bona for the
particular case of the Bona-Masso formulation~\cite{Arbona98}.  We
have also shown the effectiveness of our algorithm on a couple of
different formulations of the evolution equations.


\bibliographystyle{apsrev}
\bibliography{referencias}


\end{document}